\documentclass[twocolumn]{aastex6}	
\slugcomment{Accepted for publication, ApJL}

\usepackage{color}

\shorttitle{K2, CRTS, and ASAS-SN Observations of AR Sco}
\shortauthors{Littlefield et al.}

\begin{document}

\title{Long-term photometric variations in the candidate white-dwarf pulsar AR Scorpii from K2, CRTS, and ASAS-SN observations}

\author{Colin Littlefield,\altaffilmark{1} Peter Garnavich,\altaffilmark{1} Mark Kennedy,\altaffilmark{1, 2} Paul Callanan,\altaffilmark{2} Benjamin Shappee,\altaffilmark{3} Thomas Holoien\altaffilmark{4}}

\altaffiltext{1}{Department of Physics, University of Notre Dame, Notre Dame, IN, USA}
\altaffiltext{2}{Department of Physics, University College Cork, Cork, Ireland}
\altaffiltext{3}{Carnegie Observatories, Pasadena, CA, USA}
\altaffiltext{4}{Department of Astronomy, The Ohio State University, Columbus, Ohio, USA}

\begin{abstract}

We analyze long-cadence Kepler \textit{K2} observations of AR Sco from 2014, along with survey photometry obtained between 2005 and 2016 by the Catalina Real-Time Sky Survey and the All-Sky Automated Survey for Supernovae. The \textit{K2} data show the orbital modulation to have been fairly stable during the 78 days of observations, but we detect aperiodic deviations from the average waveform with an amplitude of $\sim 2$\%\ on a timescale of a few days. A comparison of the \textit{K2} data with the survey photometry reveals that the orbital waveform gradually changed between 2005 and 2010, with the orbital maximum shifting to earlier phases. We compare these photometric variations with proposed models of this unusual system.

\end{abstract}


\section{Introduction}

\begin{figure*}
\includegraphics[width=\textwidth]{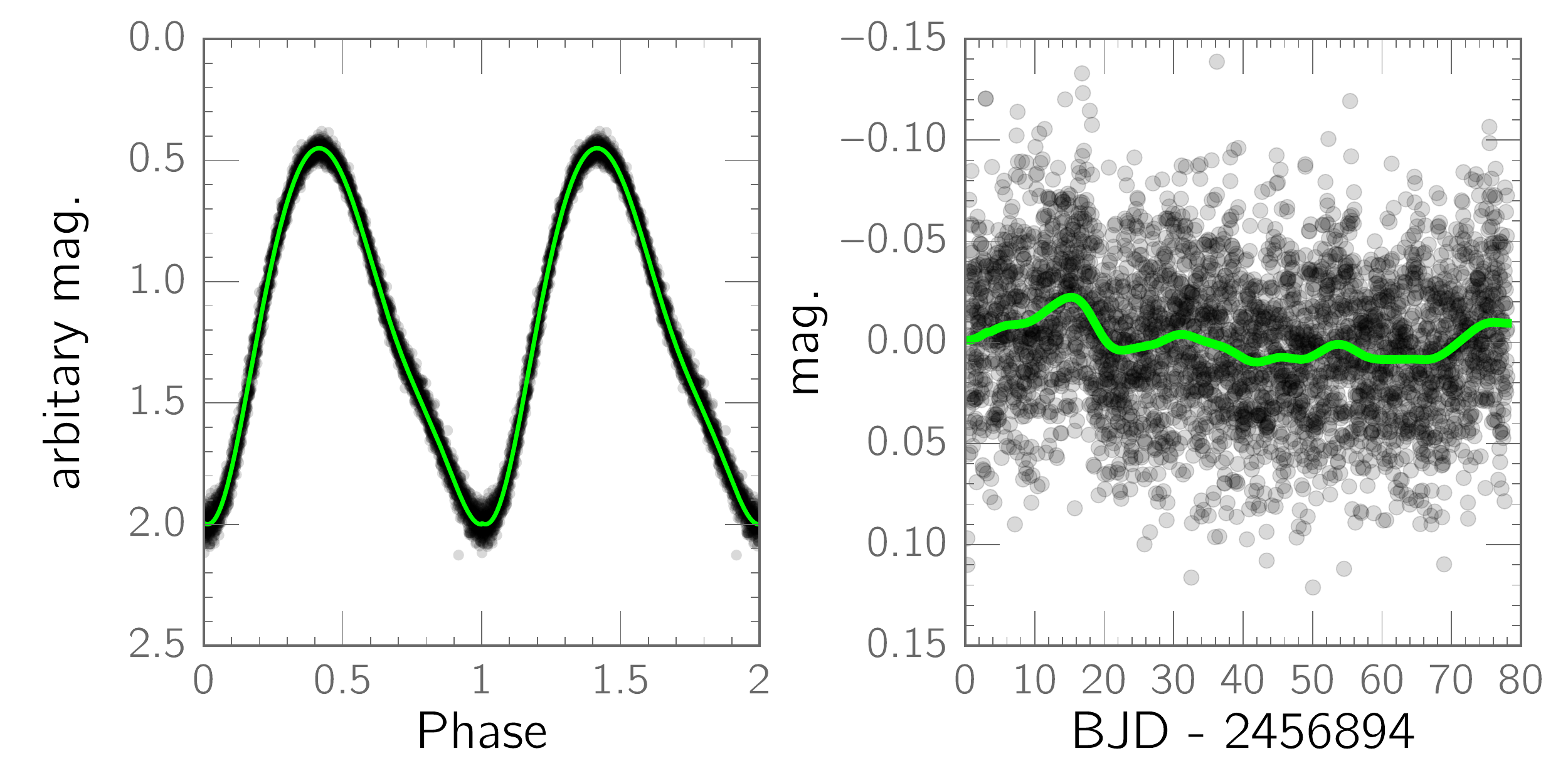}
\caption{{\it Left: } A phase plot of the \textit{K2} data, using the orbital ephemeris from \citet{marsh}. The green line is the best-fit polynomial. The data are repeated for clarity.
{\it Right:} The full \textit{K2} light curve after subtracting the polynomial model of the phase plot. The smoothed trendline highlights bumps near T~=$\sim$15~d and T~=$\sim$75~d.}
\label{phaseplot}
\end{figure*}

AR Scorpii (AR Sco) is an unprecedented binary system featuring a white dwarf (WD) that  generates highly periodic pulses across the electromagnetic spectrum every 1.97~min, even at radio wavelengths \citep{marsh}. The system's total luminosity exceeds the combined luminosity of the WD and its red dwarf companion by an average factor of $\sim$4, and its spectral energy distribution is consistent with synchrotron radiation, with the low X-ray luminosity implying minimal accretion \citep{marsh}. Interpreting AR Sco as the first WD pulsar, \citet{buckley} established that the optical pulses are highly linearly polarized and argued that the observed properties of AR Sco are consistent with a strongly magnetized, rapidly rotating WD whose spin-down powers the system's luminosity. In this scenario, the WD's magnetic axis is nearly perpendicular to its rotational axis, and the synchrotron radiation is produced when the WD's open field lines sweep across the secondary, accelerating electrons in its wind \citep{geng}. High-angular-resolution interferometric observations have shown the radio-emitting region to be smaller than $\sim0.02$~AU, implying the absence of a radio-bright outflow, such as a collimated jet \citep{marcote}.

As an alternative to the WD-pulsar model, \citet{katz} proposed two hypotheses. In the first, the WD's magnetic field sweeps over the face of the secondary, leading to the formation of a bow wave on the leading face of the secondary. Magnetic dissipation occurs preferentially in this bow wave, causing the observed orbital maximum at $\phi_{orb}\sim0.4$, where superior conjunction of the secondary occurs at $\phi_{orb} = 0.5$. In Katz's alternative hypothesis, the WD's spin axis is misaligned with the binary's orbital axis, and the WD's magnetic moment is inclined with respect to its spin axis. As a result, the magnetic field experienced by the secondary varies with orbital phase, leading to a photometric modulation at the orbital period. In this second hypothesis, the misalignment causes a precession, and Katz predicts that the phase of maximum light drifts on timescales of $\sim$20-200 years. This provides an observational test between the various models.

In the optical, the system exhibits two principal periodicities: a 3.56-hour orbital period, and the aforementioned 1.97-min, double-peaked pulsation, the amplitude of which can be as large as $\sim$1.5~mag, corresponding with the beat period between the WD spin and binary orbital periods. Curiously, the peak of the orbital modulation occurs at $\phi_{orb}\sim$ 0.4 and is therefore offset from the time of superior conjunction at $\phi_{orb}$ = 0.5, the phase at which maximum light from the irradiated inner hemisphere of the secondary would normally be observed. 

We analyze eleven years of survey photometry of AR Sco, as well as Kepler \textit{K2} observations, to investigate the long-term stability of the orbital modulation in order to provide additional constraints for theoretical explanations of the system.

\section{Data \& Analysis}
\subsection{Kepler \textit{K2} photometry}

\begin{figure}
\includegraphics[width = 0.5\textwidth]{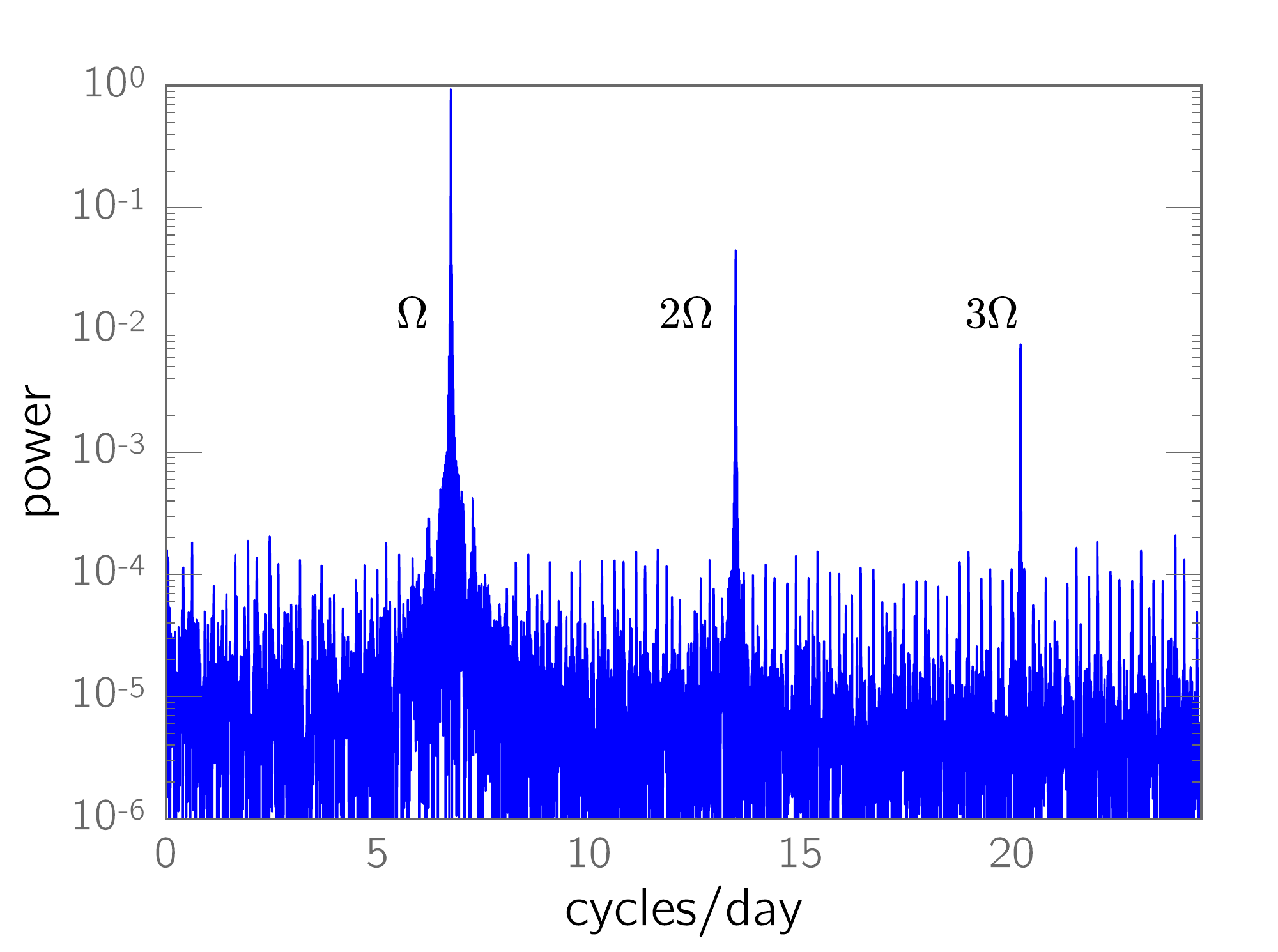}
\caption{A Lomb-Scargle power spectrum of the \textit{K2} light curve, with the orbital frequency ($\Omega$) and its first two harmonics labeled. There is no evidence of other periodicities within this frequency range. The highest frequency plotted is the Nyquist frequency.}
\label{power}
\end{figure}

Before \citet{marsh} uncovered AR Sco's extraordinary nature, the Kepler satellite observed AR Sco in long-cadence mode as part of program GO2049 (P.I. Andrej Prsa) during Campaign 2 of the \textit{K2} mission between 23 Aug. 2014 and 10 Nov. 2014. Unfortunately, the 30-minute cadence of the \textit{K2} data means that the 1.97-min pulses are not temporally resolved, but the data nevertheless provide a unique opportunity to assess the stability of the orbital waveform across the 79-day \textit{K2} run.

To extract the photometry, we downloaded the pixel file for AR Sco and performed aperture photometry on AR Sco's centroid in each image. All datapoints affected by thruster firings were removed. When phased to the orbital ephemeris in \citet{marsh}, the \textit{K2} observations show the orbital modulation to be remarkably consistent, with no obvious variations in the system's overall brightness (Figure~\ref{phaseplot}, left panel). The phase plot shows that the orbital modulation peaked near $\phi_{orb}\sim$ 0.4, as observed in \citet{marsh}. The rise to maximum is steeper than the decline to minimum, with the latter showing a change in slope near $\phi_{orb}\sim$ 0.7.

\begin{figure*}[ht!]
\includegraphics[width=\textwidth]{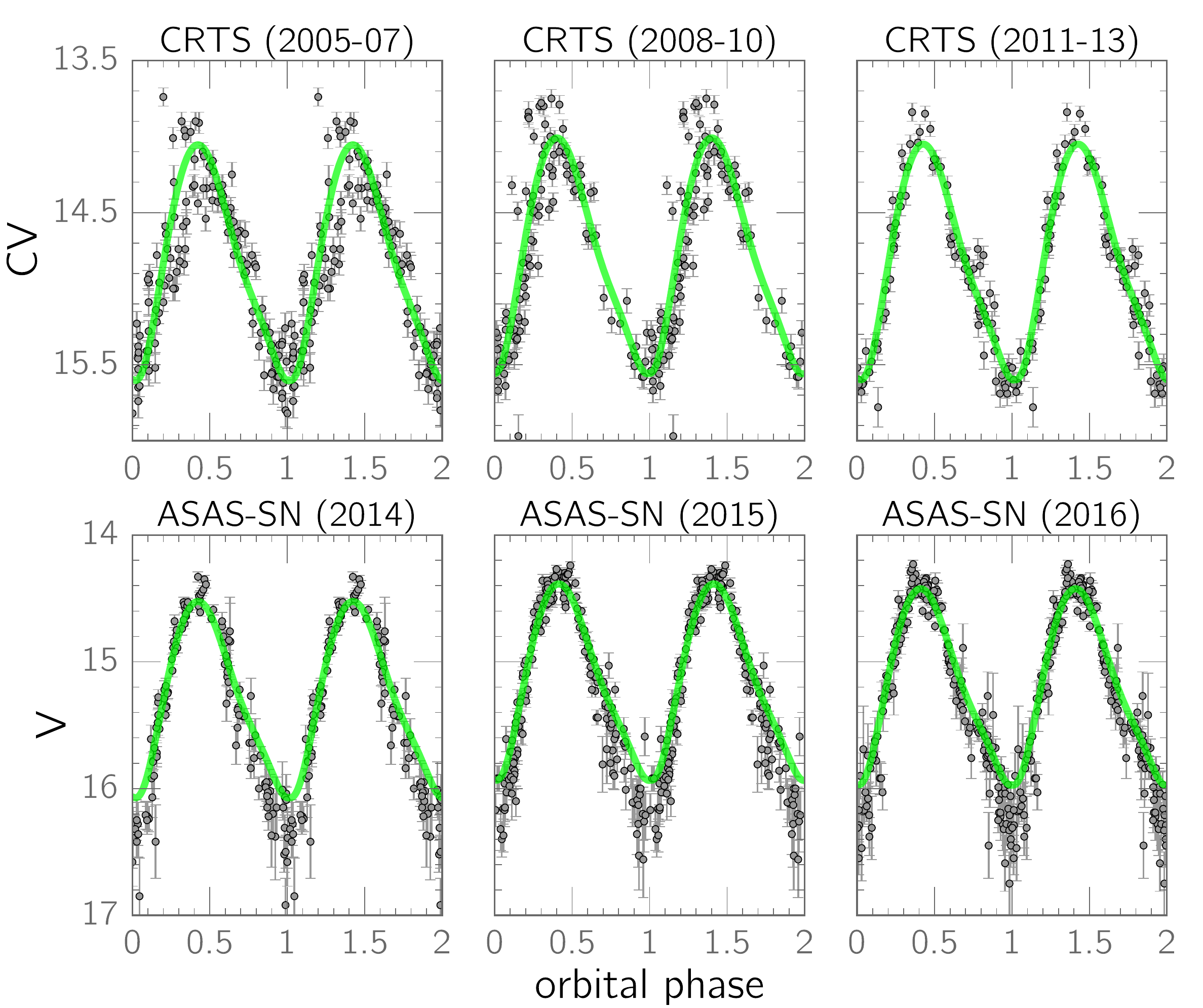}
\caption{The CRTS and ASAS-SN data. with the \textit{K2} waveform superimposed in green. The \textit{K2} waveform (green line) was fitted to each bin by applying vertical and horizontal translations to minimize $\chi^{2}.$ In the earliest CRTS bin, there is significant deviation from the \textit{K2} modulation, and the rising part of the light curve is fainter than in subsequent bins. The bright outliers in the CRTS data are consistent with beat pulses. The ``CV'' bandpass denotes unfiltered data with a $V$ zeropoint. A bandpass difference between the unfiltered \textit{K2} data and the $V$-band ASAS-SN photometry likely accounts for the underprediction of the amplitude of the ASAS-SN orbital modulation.}
\label{crts_asassn}
\end{figure*}

\begin{figure*}[ht!]
\includegraphics[width=\textwidth]{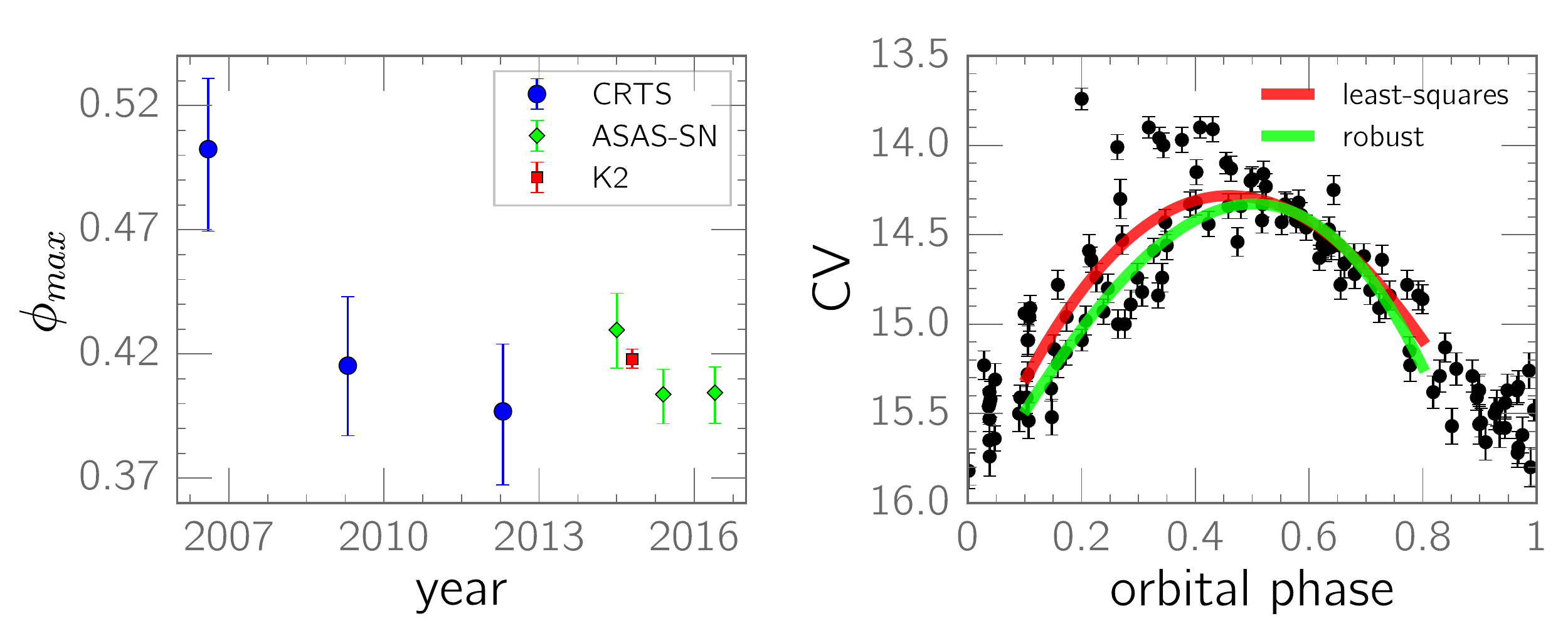}
\caption{{\it Left:} The phase of the orbital maximum in each bin, as determined by fitting outlier-resistant polynomials to the phase plots in Fig.~3. {\it Right:} The  CRTS bin showing the largest phase shift (2005-2007), with least-squares and robust polynomials superimposed. The pure least-squares fit is skewed by a handful of bright outliers from beat pulses, causing it to peak at an earlier phase than the robust polynomial. The RANSAC algorithm used to fit the robust polynomial identifies these outliers and excludes them from the fit.}
\label{maxima}
\end{figure*}

A power spectrum of the light curve (Figure~\ref{power}) contains the fundamental orbital frequency and its next two harmonics. We did not search for frequencies above the Nyquist frequency for the data. The orbital period in the \textit{K2} data is 0.148533(7)~d, consistent with the value reported by \citet{marsh}. There is no evidence of additional periodicities in the power spectrum.

Based on the phase plot, we computed a polynomial that describes the orbital modulation as a function of orbital phase. We used this polynomial to subtract the orbital modulation from the full light curve to search for subtle variations in the overall brightness, and we plot these residuals in the right panel of Figure~\ref{phaseplot}. There is a significant brightness variation approximately 15 days into the \textit{K2} run, with an amplitude of 2\%. Another possible rise is seen at the end of the run. The origin of these week-long brightness variations is unclear, but we viewed the images from the pixel file to ensure that they are not artifacts caused by the passage of asteroids through the image.

\subsection{CRTS and ASAS-SN photometry}

The excellent signal-to-noise ratio of the orbital modulation in the \textit{K2} data enables us to test whether it is consistent with the orbital modulation in the Catalina Real-Time Sky Survey \citep[CRTS;][]{drake} and the All-Sky Automated Survey for Supernovae \citep[ASAS-SN;][]{shappee, kochanek}. Together, the photometry from these two surveys provides coverage of AR Sco from 2005-2016. Because of sparse sampling, we divide the CRTS photometry into a trio of three-year bins (2005-2007, 2008-2010, and 2011-2013) and the ASAS-SN data into three one-year bins (2014, 2015, and 2016); in none of the bins were CRTS and ASAS-SN data merged. We phased the data in each bin using the ephemeris from \citet{marsh} and then fit the average \textit{K2} waveform to each bin using $\chi^{2}$ minimization to solve for two free parameters: a phase shift and a magnitude offset.

The resulting fits are shown in Fig.~\ref{crts_asassn}. Although the three CRTS-only bins generally agree with the \textit{K2} waveform, there is a striking trend: the shape of the rising part of the light curve changes with time. In the earliest data (2005-2007), a number of observations between $0.2 < \phi_{orb} < 0.4$ are significantly fainter than expected from the \textit{K2} fit, some by as much as a half-magnitude. Although there are some points that are considerably brighter than the \textit{K2} fit, these observations were probably contaminated by beat pulses, during which AR Sco can flare by a factor of four in $\lesssim$ 30 seconds \citep{marsh}. In the next bin (2008-2010), the rising part of the light curve is still fainter than the \textit{K2} fit, but the discrepancy is not as pronounced. Finally, the most recent CRTS bin (2011-2013) shows excellent agreement with the \textit{K2} light curve at all phases.

The \textit{K2} fit to the ASAS-SN data underpredicts the amplitude of the orbital variation, but this is likely the result of a difference in bandpass. The \citet{marsh} spectrum shows a strong contribution from the M-dwarf from $\sim$700-900~nm. While an unfiltered CCD is sensitive at these wavelengths, the $V$-band has negligible transmission redward of 700~nm. As a result, the contrast between the heated inner hemisphere and the presumably unheated outer hemisphere will be greater in the $V$-band ASAS-SN data than in the unfiltered \textit{K2} and CRTS data, giving rise to a larger orbital modulation at shorter wavelengths.

\floattable
\begin{deluxetable}{ccccccc}

\tablecaption{Phase of the orbital maximum}

\tablehead{\colhead{Source} & \colhead{Band\tablenotemark{$a$}}& \colhead{$\phi_{max}$ (robust)} & \colhead{sig.\tablenotemark{$b$}} & \colhead{$\phi_{max}$ (least sq.)} & \colhead{sig.\tablenotemark{$c$}} & \colhead{peak mag.\tablenotemark{$d$}}}
\startdata
CRTS (2005-2007) & CV & 0.505$^{+0.031}_{-0.028}$ & 0.000 & 0.460$^{+0.022}_{-0.022}$ & 0.000 & 14.32$\pm$0.07 \\
CRTS (2008-2010) & CV & 0.415$^{+0.028}_{-0.028}$ & 0.539 & 0.395$^{+0.022}_{-0.018}$ & 0.712 & 14.10$\pm$0.07 \\
CRTS (2011-2013) & CV & 0.396$^{+0.030}_{-0.027}$ & 0.726 & 0.398$^{+0.020}_{-0.016}$ & 0.602 & 14.01$\pm$0.09 \\
ASAS-SN (2014) & V & 0.429$^{+0.015}_{-0.015}$ & 0.022 & 0.424$^{+0.013}_{-0.014}$ & 0.012 & 14.49$\pm$0.03 \\
K2 (2014) & $K_{p}$ & 0.418$^{+0.004}_{-0.004}$ & n/a & 0.418$^{+0.003}_{-0.002}$ & n/a & n/a\\
ASAS-SN (2015) & V & 0.404$^{+0.011}_{-0.010}$ & 0.077 & 0.401$^{+0.009}_{-0.007}$ & 0.064 & 14.38$\pm$0.03 \\
ASAS-SN (2016) & V & 0.405$^{+0.012}_{-0.011}$ & 0.164 & 0.403$^{+0.010}_{-0.008}$ & 0.158 & 14.41$\pm$0.03\\
\enddata

\tablenotetext{a}{CV = unfiltered with $V$ zeropoint}
\tablenotetext{b}{Fraction of simulations of robust fits in which the simulated phase shift was larger than the observed value.}
\tablenotetext{c}{Same as $b$, but for the least-squares fits.}
\tablenotetext{d}{Average maximum magnitude of the orbital modulation in robust fits.}

\label{table}

\end{deluxetable}

\subsection{Phase of the Orbital Maximum}

We fit third-order polynomials to the orbital maximum in each of the bins to estimate the phase of maximum light. The initial fits to the data used a conventional least-squares-minimization approach, but the quality of these fits is adversely affected a number of bright outliers between orbital phases $\sim$0.1-0.4. \citet{marsh} showed that the amplitude of the beat pulsations is largest at these orbital phases, so observations contaminated by beat pulses will drag the maximum of a conventional least-squares polynomial towards earlier phases. Thus, we also used random-sample consensus \citep[RANSAC; ][]{fischler}, a machine-learning algorithm that identifies and masks outliers in a dataset, to fit a robust polynomial to each bin. RANSAC iteratively selects random subsets of a dataset, fitting them with a specified model (in this case, a third-order polynomial). The model describing the subset is compared to all points not in the subset, and if a certain number of them agree with the model to within a specified tolerance, they are considered inliers, as are the points in the original subset. The model is then fitted to the set of inliers to produce the provisional best-fit model; outliers are ignored. RANSAC repeats this process with different subsets, attempting to find the provisional best-fit model with the lowest residuals.

The left panel in Figure~\ref{maxima} plots the phases of orbital maximum as measured by the robust polynomials, and to illustrate how the choice of fitting algorithm impacts the measured phase of orbital maximum, the right panel shows a comparison of the least-squares and RANSAC fits to the first CRTS bin. Additionally, Table~\ref{table} lists the phases of maximum light using the two fitting methods, along with 1$\sigma$ uncertainties.

Both fitting techniques agree that there was a phase shift towards later phases in the 2005-2007 CRTS bin, but the robust polynomial shows it to be larger and more statistically significant. Although the contemporaneous \textit{K2} and 2014 ASAS-SN observations are weakly suggestive of a small phase shift, the data after the first CRTS bin are consistent with an essentially constant phase of orbital maximum, so any periodicity in the phase of maximum light must be much longer than the available baseline of data. In addition, as Table~\ref{table} indicates, the maximum of the orbital light curve was $\sim$0.2 mag fainter in the earliest CRTS bin than in the other two CRTS bins, and the 2014 ASAS-SN bin was $\sim$0.1 mag fainter than the 2015 and 2016 ASAS-SN data.

Because of a bandpass difference between CRTS and ASAS-SN, the peak magnitudes of the CRTS and ASAS-SN orbital modulation cannot be directly compared with each other. It is for a similar reason that we did not attempt to measure the peak magnitude in the \textit{K2} data; a single \textit{Kepler} magnitude cannot be reliably compared with magnitude estimates from other sources on the level of a tenth of a magnitude.

For the CRTS and ASAS-SN data, we used 2,000 Monte Carlo simulations to test the susceptibility of the fitting procedure to false phase shifts induced by the combined effects of (1) sparse sampling, (2) propagated uncertainties from the orbital ephemeris, and (3) the difficulty of disentangling the orbital modulation from the beat pulses. The simulations made use of 22~hours of unfiltered photometry of AR Sco obtained at a typical cadence of $\sim$5~s with the University of Notre Dame's 80-cm Sarah L. Krizmanich Telescope (SLKT) during 2016 and 2017; at this cadence, the beat pulsations are reasonably well-sampled.\footnote{The SLKT data will be analyzed in a forthcoming paper.} The Monte Carlo procedure was as follows. Each simulation began by computing new orbital phases for all data, based on the uncertainties from the \citet{marsh} orbital ephemeris. For each CRTS or ASAS-SN observation in a given bin, we found the SLKT observation with the most similar orbital phase. To simulate random sampling of the beat pulse, we then randomly selected an SLKT observation obtained within $\pm\frac{1}{2}P_{beat}$ of that point and calculated the average SLKT magnitude within a timespan equal to the exposure time of the survey photometry. Using this technique, each simulation created a unique synthetic light curve whose sampling and time resolution match that of the underlying survey photometry. Finally, each simulated light curve was fit with third-order least-squares and RANSAC polynomials. The phase of maximum light and the peak magnitude of the orbital modulation were extracted from the fits.

Since sparse sampling is not an issue with the \textit{K2} photometry, we simply simulated Gaussian scatter in the flux for each \textit{K2} observation as well as propagated uncertainties from the orbital ephemeris. While this results in comparatively small uncertainties, there is probably an unquantified systematic error stemming from the fact that the beat pulses are fully blended into the orbital modulation, thereby distorting the orbital profile. Each \textit{K2} integration includes about 15 beat pulses, and because the beat pulses are strongest before the orbital maximum, their contamination will shift the orbital maximum towards earlier phases. The uncertainty for the \textit{K2} phase of orbital maximum does not model this effect.

For both the least-squares and RANSAC fits, Table~\ref{table} lists the fraction of simulations for each bin in which the simulated phase shift was larger than the observed phase shift. The results suggest that the three aforementioned effects are insufficient to produce the observed phase shift in the earliest CRTS bin; not one of the simulated phase shifts for that bin was larger than the measured value. Additionally, the simulations suggest that the small phase shift in the 2014 ASAS-SN bin is of marginal significance, but given the comparatively small size of this phase shift relative to its 1$\sigma$ uncertainty, it is possible that our simulations did not fully account for all possible causes of false phase shifts. 

\section{Conclusion}

The \textit{K2} observations from 2014 establish limits on the stability of the optical orbital modulations of AR~Sco on timescales of months and show low-amplitude, apparently aperiodic fluctuations with an unknown source. Because it is well-defined with very little scatter, the \textit{K2} orbital modulation is a useful point of comparison for the orbital waveforms of the CRTS and ASAS-SN datasets, helping to establish that in early CRTS observations, the peak of the orbital waveform was considerably fainter than in subsequent years.

An analysis of the orbital phase of maximum light reveals a significant phase shift in the earliest CRTS data (2005-2007), but the data from 2008-2016 are consistent with an unchanged phase of orbital maximum. In the 2005-2007 CRTS data, the peak magnitude of the orbital modulation is about $\sim$0.2~mag fainter than in the remaining CRTS bins. The apparent lack of a coherent trend in the phase of orbital maximum rules out sinusoidal variations on timescales of $\sim$20 years, but with only 11 years of observations, there is insufficient data to discount the possibility of a longer periodicity. Katz's misaligned-spin model predicts a precessional period of up to several centuries, so sustained long-term monitoring of the orbital modulation will be necessary to test this possibility.

\acknowledgments

We thank the referee for a thoughtful report that led to the improvement of this paper.


\begin{thebibliography}

\bibitem[Buckley et al.(2017)]{buckley} Buckley, D. et al. 2016, Nature Ast., 1, 29
\bibitem[Drake et. al.(2009)]{drake} Drake, A.~J. et al. 2009, ApJ, 696, 870
\bibitem[Fischler \& Bolles (1981)]{fischler} Fischler, M.~A. \& Bolles, R.~C. 1981, Commun. ACM, 24, 381
\bibitem[Katz(2017)]{katz} Katz, J.~I. 2017, ApJ, 835, 150
\bibitem[Geng, Zhang, \& Huang(2016)]{geng} Geng, J.-J., Zhang, B., \& Huang, Y.-F. 2016, ApJ, 831, L10
\bibitem[Kochanek et al.(2017)]{kochanek} Kochanek et al. 2017, arXiv:1706.07060, submitted to PASP
\bibitem[Marcote et al.(2017)]{marcote} Marcote, B., Marsh, T.~R., Stanway, E.~R., Paragi, Z., \& Blanchard, J.~M. 2017, \aap, 601, L7
\bibitem[Marsh et al.(2016)]{marsh} Marsh, T.~R. et al. 2016, Nature, 537, 374
\bibitem[Shappee et al.(2014)]{shappee} Shappee, B. et al. 2014, ApJ, 788, 48

\end{thebibliography}
\end{document}